\documentclass[12pt]{article}
\usepackage{epsfig}

\input epsf
\begin{document}

\def\cA{{\cal A}}
\def\cL{{\cal L}}
\def\cD{{\cal D}}
\def\g{\gamma}
\def\ss{\scriptscriptstyle}
\def\gl{\gamma_{\scriptscriptstyle L}}
\def\gr{\gamma_{\scriptscriptstyle R}}
\def\gf{\gamma_5}
\def\d{\delta}
\def\e{\eta}
\def\diag{\hbox{diag}}
\def\pl{\partial}
\def\hf{{1\over 2}}
\def\ol#1{\overline{#1}}
\def\Dsl{\hbox{/\kern-.6700em\it D}} 
\def\dsl{\hbox{/\kern-.5300em$\partial$}}
\def\veps{\varepsilon}
\def\eps{\epsilon}
\def\ebar{\ol{\eta}}
\def\pbar{\ol{\psi}}
\def\zbar{\ol{\zeta}}
\def\xbar{\ol{\xi}}
\def\lbar{\ol{\lambda}}
\def\cbar{\ol{\chi}}
\def\eqa{\begin{eqnarray}}
\def\eeqa{\end{eqnarray}}
\def\eq{\begin{equation}}
\def\eeq{\end{equation}}

\def\nn{\nonumber}
\def\cc{\hbox{c.c.}}
\def\cL{{\cal L}}
\def\veps{\varepsilon}

\begin{titlepage}

\hfill hep-th/0409242

\hfill

\vspace{20pt}

\begin{center}
{\Large \textbf{Stability in Asymptotically AdS Spaces}}
\end{center}

\vspace{6pt}

\begin{center}
\textsl{M. Kleban $^{a}$, M. Porrati $^{b}$ and R. Rabadan $^{a}$}
\textsl{}
\vspace{20pt}

\textit{$^a$ Institute For Advanced Study, Einstein Drive, Princeton NJ 08540}

\vspace{4pt}

\textit{$^b$ Department of Physics, New York University\\ 4 Washington Pl.,
New York NY 10003}

\vspace{10pt}

\textit{}
\end{center}
\vspace{12pt}

\begin{center}
\textbf{Abstract}
\end{center}

\vspace{4pt} {\small \noindent
We discuss two types of instabilities which may arise in string theory compactified to asymptotically
AdS spaces:
perturbative, due to discrete modes in the spectrum of the Laplacian,
and non-perturbative, due to brane nucleation.
In the case of three dimensional Einstein manifolds,
we completely characterize the presence of these instabilities,
and in higher dimensions we provide a partial classification.  The
analysis may be viewed as an extension of the
Breitenlohner-Freedman bound. One interesting result is that,
apart from a very special class of exceptions, all Euclidean asymptotically AdS
spaces with more than one conformal boundary component are
unstable, if the compactification admits BPS branes or scalars
saturating the Breitenlohner-Freedman bound.  As examples, we analyze quotients of
AdS in any dimension and AdS
Taub-NUT spaces, and show a space which was previously discussed in the
context of AdS/CFT is unstable both perturbatively and
non-perturbatively. }

\vfill
\vskip 5.mm
 \hrule width 5.cm
\vskip 2.mm
{\small
\noindent e-mail: matthew@ias.edu, massimo.porrati@nyu.edu, rabadan@ias.edu
}
\end{titlepage}
\tableofcontents
\section{Introduction}

In string theory compactifications, asymptotically AdS spaces can suffer from several types of instabilities.  In this note we will discuss two of these, and present criteria for stability by analyzing the spectrum of the Euclidean Laplacian.  Several interesting results will emerge from our analysis, one being that spaces with more than one conformal boundary component are typically unstable.

In~\cite{mm04}, Maldacena and Maoz discuss some examples of Euclidean spaces with two conformal boundaries. From the point of view of the dual field theory, one might expect that the theories on the two boundaries would be uncoupled; i.e. that the correlation functions would factorize. However, because properly normalized bulk correlators are not zero when evaluated between the two boundaries,  this expectation is not consistent with the bulk theory. The mystery is then what is the holographic dual to such spaces.  Our results show that such examples are generically unstable, which may to some extent alleviate this
puzzle.\footnote{In Lorentzian asymptotically AdS spaces there are often disconnected boundary components separated by horizons \cite{gsww99}. One well-known example is the AdS black hole, where the two conformal boundaries represent the two halves of a thermofield double.  Such examples are well understood (although subtleties still exist there, see {\it e.g.} \cite{maldet}), and will not be our concern here.}

A useful quantity in our analysis will be the Yamabe invariant, which is closely related to the scalar  curvature of the conformal boundary.
In essence, the Yamabe invariant is the Euclidean action for a conformally coupled scalar on the boundary, normalized in a way we specify below.  If it is positive, none of the instabilities we discuss in this note can be present, due to a theorem of Lee \cite{lee94}.
Positivity also implies, by a theorem of Witten and Yau \cite{wy99}, that the conformal boundary has only one connected component.  Recently, this result has been extended to the case where the invariant is zero \cite{cai}.

On the other hand, if the Yamabe invariant is negative, there are two types of instabilities that may occur:

\begin{itemize}
\item perturbative: if there are scalars with mass $m^2 < - \lambda_0$, where $\lambda_0$ is the eigenvalue of the lowest normalizable eigenmode of the Euclidean Laplace operator. If the Yamabe invariant is negative, this can occur even for tachyons which would ordinarily be allowed in AdS; that is, fields satisfying the Breitenlohner-Freedman bound.  In particular, fields saturating the BF bound may be problematic.
\item non-perturbative:  if there are BPS branes, there will be an instability due to brane nucleation, because large branes have an energy which is unbounded from below as they approach the boundary.
\end{itemize}

In what follows we will discuss the specific criteria which must be satisfied to avoid these instabilities.  The presence or absence of negative $m^2$ scalars and of BPS branes is of course determined by the details of the specific compactification in question.  In this paper we will be concerned only with the spectrum of the asymptotically AdS part of the space.  With this information, and some information on the low-energy spectrum, one can check if a given compactification is stable.

The plan of the paper is as follows:  In section 2, we define the instabilities and make some general statements regarding their potential presence.

In section 3, we discuss quotients of AdS, for which we completely classify the features relevant to stability.  Since all smooth, metrically complete,
Euclidean, hyperbolic 3-manifolds are quotients of $H_3$
(Euclidean $AdS_3$), this constitutes a complete classification for three dimensions.

In section 4, we discuss some examples, and analyze the presence or absence of instabilities. One example is a one parameter family of spaces that
interpolates between stable and non-stable manifolds.  We analyze where the different instabilities appear. In particular, we show that the Bergmann space (which is a limit of this one parameter family and was discussed in \cite{bsv99}) is unstable both perturbatively and non-perturbatively.

\section{Instabilities in AdS}

In this section we define the Yamabe invariant and discuss some general properties of the spectrum of the Laplace operator in asymptotically AdS spaces. Notice than when we are studying manifolds that are not Einstein and other fields are involved, the scalar fields satisfy a more complicated equation and the analysis should be modify accordingly. For example
 \cite{mm04} constructs
several examples where extra background fields are turned on, thus significantly modifying
 the stability analysis.

We explain first how perturbative instabilities can appear; next we
discuss non-perturbative instabilities.

\subsection{The Yamabe Invariant}

The Yamabe invariant is defined for the conformal class of metrics $[h]$ on an $n$-dimensional manifold ${\cal B}$.\footnote{In this paper, the bulk manifold $\cal{M}$ has dimension $d$ and metric $g$, and $\partial \cal{M} \equiv \cal{B}$ has dimension $n = d-1$ and metric $h$.} It is the infimum of the the total scalar curvature, normalized with respect to the total volume ({\it i.e.} for metrics with unit volume), in the class $[h]$ \cite{besse}.

Let us take a metric $h$ on ${\cal B}$ with Ricci scalar $R$. By performing a conformal transformation $h \rightarrow f^{n-2 \over 4} h$, where $f: {\cal B} \rightarrow {\cal R}^+$, the Ricci scalar of $f^{n-2 \over 4} h$ is:

\eq
R \rightarrow h^{-(n+2)/(n-2)} \left( R f + 4 \frac{n-1}{n-2} \Delta (f) \right),
\eeq

Define:

\eq
Y[f] = \frac{\int_{\cal B} dx^n \sqrt{h} \left( R f^2 + 4 \frac{n-1}{n-2} (\partial f)^2 \right) }{(\int_{\cal B} dx^n \sqrt{h} f^{2n/(n-2)})^{(n-2)/n} }.
\eeq

Notice that for constant $f$, $Y$
does not change. The Yamabe invariant $\mu[h]$ is then the infimum of the $Y[f]$'s. In particular, if $R$ is positive for some metric in the conformal class then it is positive for all, and the Yamabe invariant is also positive. If there is a metric with $R = 0$ in the conformal class, the Yamabe invariant is zero, and if $R < 0$ everywhere for some metric the Yamabe invariant is negative.  For $n=2$, the Yamabe invariant is proportional to the Euler characteristic $\chi$.

Notice also that the Yamabe invariant is proportional to the action of a conformally coupled scalar.

\subsection{Spectrum of Asymptotically AdS Spaces}

All asymptotically AdS (or more precisely, asymptotically hyperbolic, since we are concerned
with Euclidean signature metrics) manifolds, with or without an Einstein metric, have a continuous spectrum for the scalar Laplacian\footnote{Here we normalize the
metric so that asymptotically it is AdS with unit radius.  For a precise definition of asymptotically
AdS or hyperbolic, see \cite{lee94}.   Roughly speaking, the condition is simply that the metric reduce to the form
$ds^2 = dz^2 + z^{-2} ds_B^2$ close to the conformal boundary, where $z \rightarrow 0$ at the boundary
and the boundary metric $ds_B^2$ is well behaved.}:
$\left[ n^2/4, \infty \right)$, as in Euclidean AdS\cite{mazzeo91}.
The gap means that tachyons with mass satisfying  $m^2 \geq -n^2/4$ are allowed, which is the famous
Breitenlohner-Freedman bound~\cite{bf}.

However, there can be discrete eigenvalues corresponding to strictly normalizable modes, $0 < \lambda <  n^2/4$. A theorem due to Lee \cite{lee94} tells us that for Einstein manifolds these modes can exist only when the Yamabe invariant of the conformal boundary is negative.  The converse has not been proven, and later in the paper we discuss a class of manifolds for which our numerical results indicate that it is violated.

\begin{figure}
\centering \epsfxsize=3in \hspace*{0in}\vspace*{.2in}
\epsffile{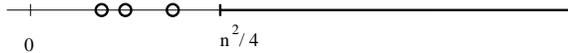}
\caption{\small Spectrum of the Laplacian for asymptotically AdS manifolds.}
\label{spectrum}
\end{figure}

\begin{figure}
\centering \epsfxsize=5in \hspace*{0in}\vspace*{.2in}
\epsffile{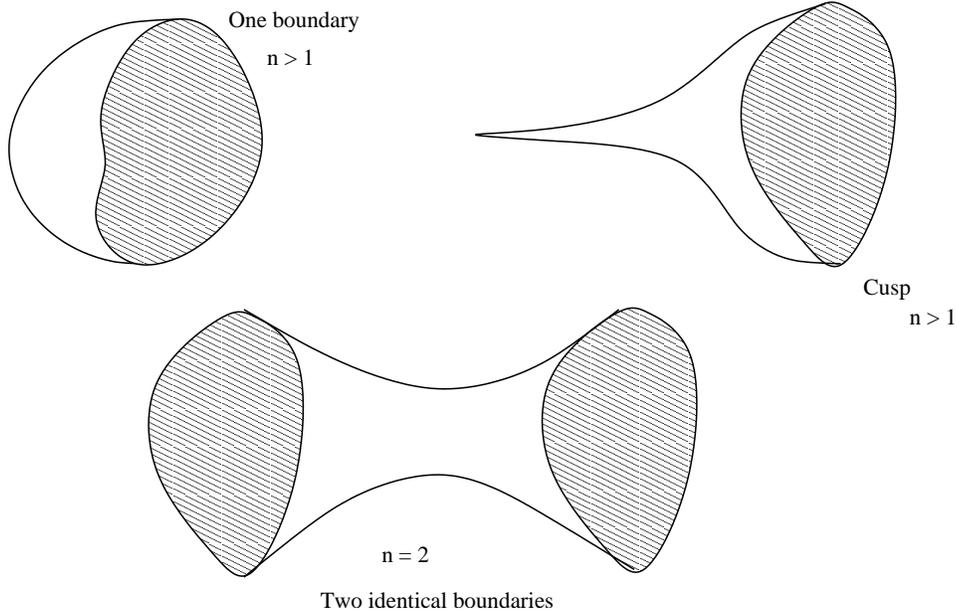}
\caption{\small All manifolds without discrete modes take on of these forms.}
\label{possible}
\end{figure}

In \cite{lw02}, Li and Wang proved that a complete manifold whose Ricci curvature is bounded from below by $Ric \geq - n$ (where $Ric$ is the lowest eigenvalue of the Ricci tensor $R_{ij}$) and $\lambda \geq n^2/4$ has  a very specific
form.\footnote{This condition is similar to but not identical with the
Strong Energy Condition in Lorentzian signature.
In dimension $n$ higher than $2$ our condition is weaker; it formally
coincides with the Strong Energy Condition for $n=2$. Another difference is
that the Strong Energy Condition is imposed on timelike vectors only,
that have no analogue in Euclidean signature.}
For asymptotically AdS spaces, we know that discrete eigenvalues, if they are present, satisfy $\lambda < n^2/4$,  so their result
can be rephrased as follows:

If there are no discrete modes and the Ricci curvature is bounded from below by its asymptotic value $Ric \geq -n$  then the manifold satisfies one of the following (see figure \ref{possible}):
\begin{itemize}
\item There is only one conformal boundary, i.e. the space is not an Euclidean wormhole.
\item The manifold has only one boundary and one cusp. The metric can be written as:
\eq
ds^2 = dt^2 + e^{2t} ds^2_N  ,
\eeq
where $N$ is a compact manifold.
\item If $d=3$, the manifold can be a ``wormhole" of the form:
\eq
ds^2 = dt^2 + \cosh{t}^2 ds^2_N .
\eeq

\end{itemize}

\subsection{Perturbative Instabilities}

By perturbative instabilities, we mean normalizable modes $\phi$ that satisfy
\eq
(- \Delta + m^2) \phi = \beta \phi,
\eeq
where $\beta < 0$ and $m$ is the mass of a scalar field.  Such wrong-sign modes would make the Euclidean path integral complex, and probably ill-defined. In other words, to ensure stability the square mass should be larger than minus the lowest eigenvalue of the Laplacian, $\lambda_0$. In the case that there are
no discrete modes, i.e. the spectrum is purely continuous,
this condition reduces to the Breitenlohner-Freedman bound.
Interesting instabilities can appear when there are discrete normalizable modes. For Einstein manifolds, that can only happen when the Yamabe invariant is negative. Therefore, there will be no instabilities in the bulk when the Euclidean action for a conformally coupled scalar on the boundary is non-negative.

The theorem of Li and Wang mentioned in the previous section
tells us that if an asymptotically AdS space has its Ricci curvature bounded from
below by its value at infinity, and if there is a scalar saturating the BF bound,
then the space can only be stable if either:

-  it has only one conformal boundary,

-  it has one boundary and one cusp,

-  it is three dimensional and is a wormhole with two identical
boundaries.\\
Note that the converse to this theorem is false
(see section 4 for an example).

The presence of a discrete mode with eigenvalue $\lambda$ causes an instability when there is a scalar with a negative mass satisfying $m^2 < - \lambda$. Although such scalars are often present from the KK reduction on the compact space, this can be avoided in some cases by taking free orbifolds that project out these modes \cite{mm04}. One extreme case where these fields cannot be
projected out is when they form a continuous spectrum.  This can happen {\it e.g.} when the conformal boundary degenerates \cite{tr00,bsv99}.

\subsection{Non-Perturbative Instabilities}

By non-perturbative instabilities, we mean nucleation of
codimension-one branes in the bulk \cite{mms98,sw99,wy99}.
The action for a Euclidean brane has a positive contribution from its tension (which is proportional to the area $A$ of the brane) and a negative contribution from the antisymmetric tensor coupling (proportional the the volume $V$ enclosed by the brane, and to the charge $q$):
\eq
S = A - q V.
\eeq

We will determine when instabilities appear, and compute the subleading corrections to the action, which will be relevant for the cases where the boundary at infinity is flat and the brane is BPS. The metric close to the boundary $\rho = \infty$ is:

\eq
ds^2 = d\rho^2 + \frac{e^{2 \rho}}{4} h_{ij} dx^i dx^j - P_{ij} dx^i dx^j + e^{-2 \rho} \alpha_{ij} dx^i dx^j + O(e^{-4 \rho}),
\eeq
where $h_{ij}$ is the metric of the boundary. $P_{ij}$ is a tensor that can be obtained by Einstein's equations from the metric $g_{ij}$ in dimensions $n = d-1 >2$:

\eq
P_{ij} = \frac{2 (n-1) R_{ij} - g_{ij} R}{2 (n-1) (n-2)}.
\eeq
For $n=2$, the trace of $P_{ij}$ can be related to the curvature of $h_{ij}$:

\eq
h^{ij} P_{ij} = \frac{R}{2}.
\eeq
The asymptotic value of the area of an $n-1$-brane close to the boundary is:

\eq
A = \frac{e^{n \rho}}{2^n} \int d^n x  \sqrt{h} \left[ 1 + e^{-2\rho} X + e^{-4 \rho} Y + O(e^{-6 \rho}) \right] + k_A,
\eeq
where $X = -2 h^{ij} P_{ij} = - R/ (n-1)$, $Y = 2 (\alpha_i^i + P_{ij}P^{ij})$ and $k_A$ is the regularized area (i.e. the term remaining after subtracting
all terms that diverge when $\rho$ goes to infinity).

The volume enclosed by this brane is:

\eq
V = \frac{e^{n \rho}}{2^n} \int d^n x \sqrt{h} \left[ \frac{1}{n} + \frac{e^{-2\rho}}{n-2} X + \frac{e^{-4 \rho}}{n-4} Y + O(e^{-6 \rho}) \right] + k_V,
\eeq
for $n$ odd.  For $n$ even, the term with a zero in the denominator is linear in $\rho$. Such terms proportional to $\rho$ give the conformal anomaly, which occurs only for an even-dimensional boundary. The terms in the expansion in
powers of $\exp(-\rho)$ are then just the anomaly polynomials in the boundary curvature. $k_V$ is an integration constant that can be understood as the regularized volume, i.e. the finite part that remains after subtracting all the divergent terms in the $\exp(-\rho)$ expansion and taking the limit $\rho \rightarrow \infty$.

The action for a brane with charge $q$ and tension $T$ is

\begin{eqnarray}
\label{lag}
S & \sim & T (A - q n V) \nonumber \\
& = & T \frac{e^{n \rho}}{2^n} \int d^n x \sqrt{h} \Big[ (1-q) \nonumber
 \\
&&  + e^{-2\rho} (1- \frac{nq}{n-2}) X + e^{-4 \rho}  (1- \frac{nq}{n-4}) Y + O(e^{-6 \rho}) \Big] + T (k_A -q n k_V) .
\end{eqnarray}

Notice that for non-BPS branes ($q < 1$) the first term dominates, and the action will diverge at the boundary. In the case of BPS branes ($q = 1$) the first term cancels, and the dominant contribution comes from the next term. The action will grow if the curvature is positive, but there will be a non-perturbative instability if the curvature of the boundary is negative. The second term will vanish if the curvature of the boundary is zero, in which case the third term is important.

For dimension $n<4$, the constant terms $k_V$ and $k_A$ are important for the stability analysis. For instance, for $n=2$,  the Lagrangian
has the following form:

\eq
L  =   T \frac{e^{2 \rho}}{4} \sqrt{g} \left\{ (1-q)  + e^{-2\rho} \left[(1 - 2 q \rho) X + \frac{4}{\sqrt{g}} (k_A -q n k_V)\right] \right\} + O(e^{-2 \rho})).
\eeq

If the brane is BPS, the large $\rho$ asymptotics is dominated by the $R \rho$ term. But if the boundary is flat the dominant term will be the constant piece that depends of the value of the metric in the whole space.  One nice example of this behavior is the BTZ black hole:

\eq
ds^2 = d\rho^2 + \sinh^2{\rho}  d\tau^2 + \cosh^2{\rho}  d\theta^2.
\eeq

The area is $A = \frac{1}{4} (e^{2\rho} - e^{-2\rho})$ and the volume $V = \frac{1}{8} (e^{2\rho} - 2 + e^{-2\rho})$. In this case
$k_A = 0 $ and $k_V = -1/4$. The action of a BPS brane is $S = T (1 - e^{-2 \rho})$ and the behavior at the boundary is determined by the regularized volume.

In three dimensions something similar happens: when the brane is BPS and the boundary has zero curvature, the asymptotically dominant term will be determined by the regularized volume and area. In four dimensions there is a term proportional to $- \rho Y$ that will control the asymptotic of the action.  In higher dimensions, similar effects can occur if enough of the boundary curvature invariants appearing in the action vanish.

To summarize, the non-perturbative instabilities can occur only if there are BPS branes, and then always when the boundary curvature is negative, and sometimes when it is zero (depending on subleading terms).

\section{Quotients of AdS}

Our first set of examples is infinite-volume, complete, smooth
hyperbolic Euclidean 3-manifolds; i.e.
Einstein manifolds with constant negative curvature. These manifolds are
quotients of the hyperbolic three space $H^3$ by a subgroup $\Gamma$ of the isometry group $SL(2,C)$. The conformal boundary of these spaces is a set of Riemann surfaces, connected through the interior.

We are interested in the existence of a discrete mode of the Laplacian with
$\lambda_0 < 1 = n^2/4$  for $n = d-1 = 2$. The existence of this mode is related to the
Hausdorff dimension of the limit set of $\Gamma$, which is defined in the following way:
select a point in the interior of the space, then the limit set is the set of
accumulation points of the orbit of this point on the conformal boundary.
It is independent of the point we started with.

According to a theorem due to Sullivan, in quotients of hyperbolic space
there will be discrete eigenmodes of the Laplacian iff the Hausdorff
dimension $d$ of the limit set is larger than $n/2$ \cite{sullivan}.
The lowest eigenvalue of the Laplacian will be $\lambda_0 = d (n - d)$.
For the case of 3-manifolds $n=2$, so the instability appears when $d > 1$. When we have more
than one boundary, it is the limit set that separates the different boundary components. Therefore
we expect that the limit set will be codimension 1 topologically, and hence $d \geq 1$ for $n=2$.
This agrees with the theorem of Li and Wang:
if there is one boundary (or a boundary and a cusp)
the space might not have discrete modes. The borderline case is when the conformal boundary
is two identical (both topologically and geometrically) Riemann surfaces.
The limit set is a circle, and the group $\Gamma$, which has a geometric circle as its limit set,
is called Fuchsian.  This is the case considered by Maldacena and Maoz in \cite{mm04},
and is the only space with two boundaries without discrete modes.

So, only very specific cases with several boundaries have no discrete
modes.  Notice that for higher dimensional quotients of AdS, if the boundary has more than one component, the limit set will have dimension $d \geq n-1$.
That means that quotients of AdS with several boundaries will always have
discrete modes in their spectrum for $n>2$, which is again consistent with the
theorem of Li and Wang.

We can summarize this section by saying that in any dimension,
quotients of AdS with more than one connected boundary component
have discrete modes, except for the case of two identical boundaries in
AdS$_3$ with metric:

\eq
ds^2 = dt^2 + \cosh{t}^2 ds^2_N  .
\eeq

\section{A Family of Examples in Four Dimensions}

In this section we exemplify our general analysis by investigating the
stability of a one-parameter family of 4-manifolds.  Qualitatively, these are obtained by squashing the boundary 3-sphere and then filling it with a hyperbolic metric.  In the limit of infinite squashing, the space becomes the Bergmann space analyzed in \cite{bsv99}.

\subsection{The Conformal Boundary: Negative Curvature Spheres}

To begin with, parametrize a three dimensional sphere by the coordinates $(\theta,\phi,\psi)$ with ranges: $0 \leq \theta \leq \pi$, $0 \leq \phi \leq 2 \pi$ and  $0 \leq \psi \leq 4 \pi$.

\begin{figure}
\centering \epsfxsize=3in \hspace*{0in}\vspace*{.2in}
\epsffile{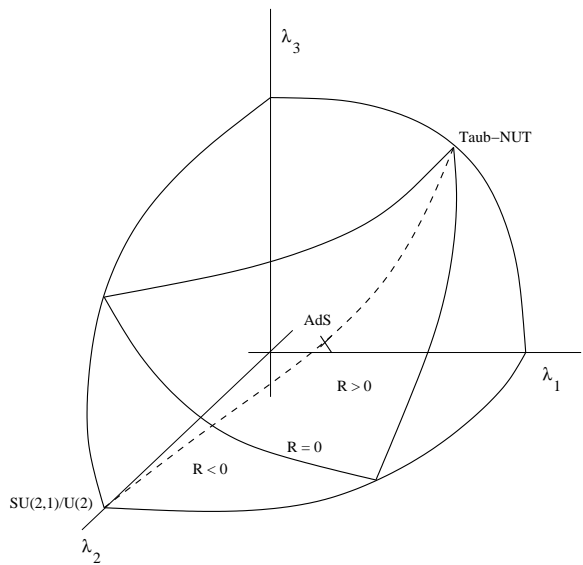}
\caption{\small Parametrization of constant curvature three spheres by $\vec{n}$.}
\label{quadrant}
\end{figure}

We are interested in metrics of constant scalar curvature on the 3-sphere.  Surprisingly, unlike the case of the 2-sphere, the curvature can take any real
value.   The metrics can be parametrized by a normal vector with positive components: $(\lambda_1, \lambda_2, \lambda_3) / \sqrt{(\lambda_1^2+ \lambda_2^2+ \lambda_3^2)}$ (see figure \ref{quadrant}).

The metric is then:
\eq
ds^2 =  \lambda_1^2 \sigma_1^2 + \lambda_2^2 \sigma_2^2 + \lambda_3^2 \sigma_3^2 ,
\eeq
where the $\sigma_i$ are the usual $SU(2)$ one-forms:
\eqa
\sigma_1 & = & \frac{1}{2} (\cos{\psi} d\theta + \sin{\psi}
\sin{\theta} d\phi), \nonumber \\
\sigma_2 & = & \frac{1}{2} (- \sin{\psi} d\theta + \cos{\psi} \sin{\theta}
d\phi) \nonumber, \\
\sigma_3 & = & \frac{1}{2} ( d\psi + \cos{\theta} d\phi) .
\eeqa

The Ricci scalar is:

\eq
R = 2 \frac{(\lambda_1 + \lambda_2 + \lambda_3)(-\lambda_1 + \lambda_2 + \lambda_3)(\lambda_1 - \lambda_2 + \lambda_3)(\lambda_1 + \lambda_2 - \lambda_3)}{\lambda_1 \lambda_2 \lambda_3}.
\eeq

\vspace{0.5cm}
We can simplify the structure by considering a line $\vec{\lambda}= (\lambda,1,1)$ that captures all the interesting cases. The metric is:

\eq
ds^2 =  \sigma_1^2 + \sigma_2^2 +  \lambda^2 \sigma_3^2 = \frac{1}{4} \left[ d\theta^2 + \sin{\theta}^2 d\phi^2 + \lambda^2 ( d\psi + \cos{\theta} d\phi)^2 \right].
\eeq

We will refer to these spaces as $\lambda$-spheres. We can understand them as a circle (parametrized by $\psi$) fibering a two dimensional sphere, where $\lambda$ is related to the length $L$ of the circle fibration by $L = 4 \pi \lambda$. The Ricci scalar is $R = 2 (4 - \lambda^2)$. Notice that when the circle becomes large ($\lambda > 2$) the Ricci scalar is negative, at $\lambda = 2$ the Ricci scalar is zero, and for $\lambda < 2$ it is positive. The ordinary round metric is $\lambda = 1$.  Spheres with $\lambda < 1$ are called oblate and with $\lambda > 1$ prolate.
These are not Einstein metrics for $\lambda \neq 1$, as can be seen from the Einstein tensor:

\eq
G_j^i = -
\left[
{\begin{array}{ccc}
\lambda ^{2} & 0 & 0 \\
0 & \lambda ^{2} & 0 \\
0 &  - 4\,(\lambda ^{2} - 1)\,\mathrm{cos}(\theta ) &  - 3\,
\lambda ^{2} + 4
\end{array}}
 \right] .
\eeq
\\

\subsection{AdS Taub-NUT Metrics}

To obtain a four dimensional Einstein manifold, we can fill the $\lambda$-spheres with a metric with negative cosmological constant
$\Lambda = - 3 k^2$.\footnote{In the rest of this section, the curvature
radius of the asymptotic space is $k$, not 1, unless otherwise stated.}
These spaces have been studied in \cite{cejm98,hhp98,ejm99,z03}. For $\lambda^2 < (2/3 - \sqrt{3}/3)$ we can fill the squashed sphere with either a Taub-NUT or a Taub-bolt type metric, otherwise only with Taub-NUT. Since we are interested in possible instabilities, we will study only zero or negative curvature spheres, and therefore consider only the Taub-NUT type metrics\footnote{There are several names for these metrics: AdS Taub-NUT, Quaternionic Taub-NUT, and Pedersen metrics.}. These metrics depend on a parameter $\mu$ that is related to the $\lambda$ of the sphere at infinity by $\lambda^2 = 1/(1- \mu/k^2)$:

\eq
ds^2 = \frac{4}{(1- k^2 r^2)^2}
\left[
\frac{1 - \mu r^2}{1- k^2 \mu r^4} dr^2 + r^2 (1 - \mu r^2) (\sigma_1^2 + \sigma_2^2) + r^2 \frac{1- k^2 \mu r^4}{1 - \mu r^2} \sigma_3^2
\right] . \label{t-nut}
\eeq
The radial coordinate has a range $0 \leq r \leq 1/k$. The parameter $\mu$ has units of mass square, and $k$ of mass.
The solution is regular for $-\infty \leq \mu \leq k^2$.
For generic values of the parameters, the isometry group is $SU(2) \times U(1)$, but at some particular values it is enhanced.

\begin{figure}
\centering \epsfxsize=3in \hspace*{0in}\vspace*{.2in}
\epsffile{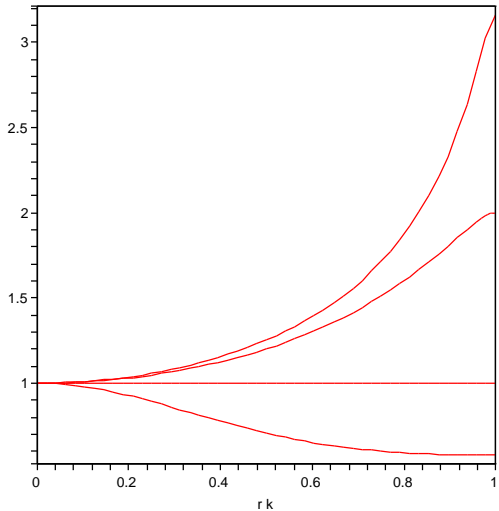}
\caption{\small Running of the $\lambda$ parameter with the radial coordinate. The lines in descending order correspond to $\lambda > 2$ (negative curvature sphere), $\lambda = 2$ (zero curvature sphere), $\lambda = 1$ (AdS case)
and  $\lambda < 1$ (the oblate case).}
\label{radial}
\end{figure}

This metric can be thought of as a series of $\lambda$-spherical shells with a $\lambda$ that runs with the radial variable:
\eq
\lambda^2(r) =  \frac{1- k^2 \mu r^4}{(1 - \mu r^2)^2},
\eeq
where $\lambda(0) = 1$ to avoid a conical defect.
In figure \ref{radial} we plot this running for several boundary values of $\lambda$.

Now we are going to consider particular values of the squeezing parameter $\lambda$ and the cosmological constant $k$.

\subsection{Taub-NUT}

In the limit that the cosmological constant vanishes ($k = 0$), we must
take take $\mu \leq 0$ to have a non-singular solution. This is just the Taub-NUT metric:

\eq
ds^2 = 4
(1 - \mu r^2) [dr^2 + r^2 (\sigma_1^2 + \sigma_2^2) ] + \frac{r^2}{1 - \mu r^2} \sigma_3^2 .
\eeq

Notice that at the boundary the circle has a finite size and the 2-sphere grows. So the boundary effectively has $\lambda \rightarrow 0$.

\subsection{AdS$_4$}

If we take a round sphere ($\lambda =1$ or $\mu = 0$) at infinity we obtain $AdS_4$:

\eq
ds^2 = \frac{4}{(1- k^2 r^2)^2}
\left[
dr^2 + r^2  (\sigma_1^2 + \sigma_2^2 + \sigma_3^2)
\right] .
\eeq

\subsection{Bergmann Space}

The limit $\mu = k^2$ ($\lambda \rightarrow \infty$ or infinite negative curvature) has the Bergmann metric:

\eq
ds^2 = \frac{4}{(1- k^2 r^2)^2}
\left[
\frac{1}{1+ k^2 r^2} dr^2 + r^2 (1 - k^2 r^2) (\sigma_1^2 + \sigma_2^2) + r^2 (1 + k^2 r^2) \sigma_3^2
\right] .
\eeq

The isometry group of this space is $SU(2,1)$. This space can be described as a coset space $SU(2,1)/U(2)$. It has been analyzed in some detail in \cite{bsv99}.  As we argue below, this space is unstable both perturbatively and non-perturbatively.

Close to the boundary, the circle grows much faster than the two dimensional sphere. By rescaling the metric we see that the conformal metric is a circle. This is a particular case where the conformal boundary has codimension larger
than one. Several examples of this sort have been studied in \cite{tr00}. In these cases the
analysis of perturbative instabilities will be similar to the Bergmann Space. The continuous
spectrum will extend,
producing unstable modes even for scalars satisfying the
Breitenlohner-Freedman bound (see section 4.8).

\subsection{Field Theory Viewpoint}

What is the interpretation of the dual field theory of these AdS Taub-NUT
spaces? Naively, by continuing the NUT charge and taking $\psi = i t$,
one gets a metric\footnote{See, for instance, \cite{amr04}.}:

\eq
ds^2 = \frac{1}{4}[d\theta^2 + \sin{\theta}^2 d\phi^2 - \lambda^2 ( dt + \cos{\theta} d\phi)^2 ] ,
\eeq
with closed time-like curves (for instance, $\psi = 0$ and $\theta < \pi/4$ or $\theta > 3 \pi/4$). To make the situation worse, if one does not impose periodicity in $t$ there are Misner singularities at the poles ($\theta=0, \pi$). To avoid them one has to make $t$ periodic, which generates closed time-like curves everywhere in the space.

The field theory in this space is probably ill-defined, which could be the origin of the strange negative entropy result of \cite{ejm99}. Perhaps the best interpretation of the action is not as a conventional free energy. A non-trivial fibration in time might be interpreted in the Hamiltonian formalism as
inserting an operator into the Boltzmann trace over states.
One very well-known example of this is the Klein bottle amplitude in string theory: the correct interpretation is a partition function with the insertion of the parity reversal operator $\Omega$.

\subsection{Non-Perturbative Instabilities}

Close to the boundary, the action for the 2-brane
has the following expansion (see eq. \ref{lag}):
\eq
S = (3-q) \left[\frac{4 \pi^2}{3} (1 - \mu) \frac{1}{\epsilon^3} +
2 \pi^2 (1-3 \mu)\frac{1}{\epsilon^2}\right] + 2 \pi^2 \frac{q + 4q \mu - 5 q \mu^2 -16 \mu + 15 \mu^2}{1-\mu} \frac{1}{\epsilon} + O(\epsilon^0),
\eeq
where we have set $k=1$.

\begin{figure}
\centering \epsfxsize=3.0in \hspace*{0in}\vspace*{.2in}
\epsffile{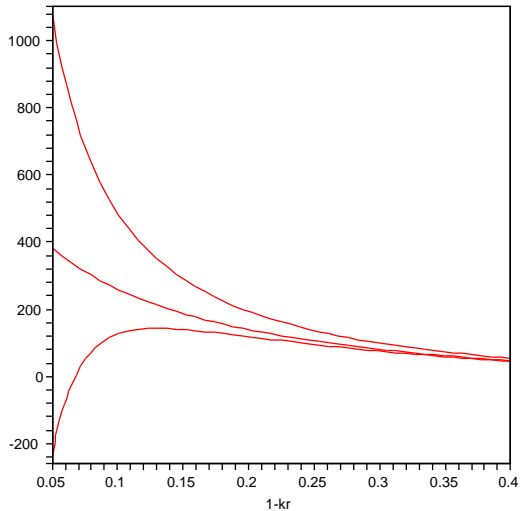}
\caption{\small Euclidean action for different boundary $\lambda$-spheres. The lines in ascending order correspond to $\lambda > 2$ (negative curvature
sphere), $\lambda = 2$ (zero curvature sphere) and $\lambda = 1$ (AdS case).}
\label{action}
\end{figure}

By saturating the BPS bound $q=3$, we can eliminate the first two terms
(see figure~\ref{action}). The divergent term near the boundary is:

\eq
S = 2 \pi^2 \frac{3 -  4 \mu}{\epsilon (1-\mu)}  + O(\epsilon^0) .
\eeq

This term is proportional to the curvature $S \sim 4 \pi^2 R (1-\mu)$, which means that spherical branes can be nucleated when the curvature of the boundary is negative, i.e. for $\lambda >2$.

\subsection{Perturbative Stability of the AdS Taub-NUT Metrics}

In this section, we analyze the perturbative stability of the AdS Taub-NUT
metrics.
For the case $\lambda \leq 2$, the Yamabe invariant is non-negative, and therefore Lee's
theorem \cite{lee94} ensures that these spaces are perturbatively stable.  Restricting
our attention to the cases $\lambda > 2$, where the boundary sphere is negatively curved,
we need to check for the existence of normalizable modes of the scalar Laplacian with eigenvalue
$\beta < 9 k^2/4$.  If such modes exist, scalars saturating the Breitenlohner-Freedman bound will make the compactification unstable.

We do not have an analytic form for the spectrum; however, by studying the behavior of
the mode equation numerically we will demonstrate that such modes do exist.  Interestingly,
they seem not to appear immediately as $\lambda$ becomes larger than 2.  At some $\lambda_* > 2$, the first bound state appears.  As $\lambda$ increases, more
and more modes appear, but the lowest eigenvalue is always bounded below by 2.
In the limit $\lambda \rightarrow \infty$ (the Bergmann space), there is a
continuum of modes with eigenvalues $\beta > 2$.

To demonstrate the existence of unstable modes, we can restrict ourselves to studying the
radial part of the equation:

\eq
-{1 \over \sqrt g} \partial_r \left( \sqrt{g} g^{rr} \partial_r \phi \right) = \beta \phi. \label{rad-la}
\eeq
We can put the equation into Schr\"odinger form, and then check for the
existence of bound states using the Bohr-Sommerfeld quantization condition.
Define
\eq
\chi = (g g^{rr})^{1/4} \phi,
\eeq
and
\eq
d \rho = dr \sqrt{g_{rr}}.
\eeq
In terms of $\chi$ and $y \equiv \sqrt{g g^{rr}}$, the equation takes the form
\eq
-\chi'' - {1 \over \sqrt{y}} \partial_\rho  \left[ y \, \partial_\rho \, \left( {1 \over \sqrt{y}} \right) \right] \chi = \beta \chi,
\label{pot}
\eeq
where
\eq
\int \chi^* \chi d \rho = 1.
\eeq

Set $k=1$ for convenience; then, for an asymptotically AdS space,
the potential $V(\rho) = - {1 \over \sqrt{y}} \partial_\rho  \left[ y \, \partial_\rho \, \left( {1 \over \sqrt{y}}
\right) \right]$ tends to a constant $V_0 = 9/4$ at infinity, which is consistent with continuity of the spectrum (see figure \ref{spectrum}).
There is a minimum when $\lambda > 2$,
and therefore bound states (and possible instabilities) will appear at some point, when the potential well is deep and broad enough.  The number of bound states $N$ is given approximately by the Bohr-Sommerfeld quantization condition:
\eq
N = {1 \over \pi} \int p dq = {1 \over \pi} \int \sqrt{V - n^2/4}d\rho ,
\eeq
where the integral is taken between classical turning points for a particle with energy $n^2/4$.  This integral can be performed numerically, and
--as we discuss below-- it diverges as the curvature of the
boundary spheres becomes more and more negative
($\mu \rightarrow 1$, recall that here $k=1$).  Unfortunately, this is not a reliable technique for estimating when the first bound state appears, since the Bohr-Sommerfeld condition is not accurate for shallow wells.  However, from studying the potential [eq.~(\ref{pot})] it seems that the first bound state does not appear as soon as the boundary curvature becomes negative, but rather somewhat later\footnote{We would like to thank Liat Maoz for discussions on this point; see also \cite{lee94}.}.  

It is clear that in the limit $\mu = 1$ (which is the Bergmann space $SU(2,1)/U(2)$ considered in \cite{bsv99}), $N \rightarrow \infty$, and so there is
an infinite number of ``bound states," starting at $\beta = 2$.  In other words, the continuum begins at $2$, rather than at $9/4$.  This can be seen from
eq.~(\ref{pot}), where we express the potential
in terms of $r$ rather than $\rho$ for convenience:
\eq
V(r, k= 1, \mu = 1) = {1 \over 16} \left( 27 + {3 \over r^2} + {4 \over 1 + r^2} \right).
\eeq
Restoring the dimensions, $V \rightarrow 2 k^2$ at the boundary, rather than
$9 k^2/4$, as it would be the case for spaces that are asymptotically AdS$_4$.

This can be seen even more directly by computing the radial Laplacian
eq.~(\ref{rad-la}) near the boundary $r=k^{-2}$.
By using eq.~(\ref{t-nut}) we get, for $k^2<\mu$,
\eq
-{1 \over \sqrt g} \partial_r \left( \sqrt{g} g^{rr} \partial_r \phi \right)
\sim -(1-kr)^4\partial_r (1-kr)^{-2}\partial_r \phi = \beta \phi,
\eeq
hence
\eq
\phi \sim (1-kr)^\alpha, \qquad
\alpha= {3\over 2} \pm \sqrt{{9\over 4} -{\beta\over k^2}},
\eeq
and the continuum lies in the range $\beta > 9k^2/4$.
At $k^2=\mu$, instead, the Laplacian is
\eq
-{1 \over \sqrt g} \partial_r \left( \sqrt{g} g^{rr} \partial_r \phi \right)
\sim -2(1-kr)^3\partial_r (1-kr)^{-1}\partial_r \phi= \beta \phi,
\eeq
so $\alpha = 1\pm \sqrt{1-\beta/2k^2}$, and the continuum begins at
$\beta=2k^2$.

Therefore, any theory with scalars of mass $m^2 < -2 k^2$ 
will be perturbatively unstable on the Bergmann space.  
For 11d supergravity, the KK reduction on S$_7$ gives 112 scalars with mass
$m^2 = -9 k^2/4$,\footnote{See~\cite{dn} eq.~(3.1.9) and (7.2.5) and Table 9;
to match notations, set their parameter $m=k/2$.}
and therefore the Bergmann space is unstable as a solution to M theory
compactified on S$_7$.

Notice that the consistent truncation of 11d supergravity to 4d, $SO(8)$ 
gauged supergravity~\cite{dwn} is stable, since all 70 scalars it contains 
obey $m^2\geq -2k^2$.
From the point of view of the boundary CFT, 4d gauged supergravity describes
only a subsector: a ring of operators closed under the OPE. Generically, 
this ring is not large
enough to describe a local CFT, so the stability of the truncation is not 
sufficient to guarantee the existence of a stable local CFT.  

The instability result we obtained for the complete KK reduction on the 
Bergmann space is robust; it is due to the fact that the family of
spaces in eq.~(\ref{t-nut}) is Einstein ($R_{\mu\nu}=-3k^2g_{\mu\nu}$), with
the {\em same} value of the cosmological constant for all $\mu$. So, since as
$\mu- k^2\rightarrow 0^-$ more and more normalizable states appear above the
Breitenlohner-Freedman bound ($m^2=-9k^2/4$), any theory containing such a
scalar is unstable, and so is the limit.\footnote{
There is a factor of 2 mismatch between our analysis and the
stability analysis of~\cite{bsv99}. In that paper, the most negative
square mass in the scalar sector is $-9k^2/8$ (to compare their normalization with
ours, set $k=\sqrt{2}$). This seems to be in
contradiction with standard KK results~\cite{dn}.}

\subsection*{Acknowledgments}
We would like to thank Stefan Bauer, Ruth Britto, Richard Canary, David Gabai, Ezra Getzler,  Juan Maldacena, Liat Maoz, and Edward Witten for helpful discussions. R.R. is supported by DOE grant DE-FG02-90ER40542; M.P. is supported in
part by NSF grant PHY-0245068; M.K. is supported by NSF grant PHY-0070928. For hospitality during the final stages of this work, the authors would
like to thank the Aspen Center for Physics, and one of us (M.K.) would
like to thank the Korea Institute for Advanced Study.

\end{document}